\documentclass[fleqn,usenatbib,useAMS]{mnras}

\usepackage{graphicx}	
\usepackage{amsmath}	
\usepackage{amssymb}	
\usepackage{multicol}        
\usepackage{bm}		
\usepackage{subfig}
\usepackage{pdflscape}	

\newcommand{\SEM}[1]{\textbf{\color{cyan} #1}}

\usepackage[T1]{fontenc}
\usepackage{ae,aecompl}

\usepackage{newtxtext,newtxmath}

\title[MAXI J1631-472]{Radio flaring and dual radio loud/quiet behaviour in the new candidate black hole X-ray binary MAXI J1631-472}
\author[Monageng et al.]
  {I.M. Monageng$^{1,3}$\thanks{E-mail: itu@saao.ac.za}, 
  S.E. Motta$^{2,3}$,
  R. Fender$^{2,4}$,
  W. Yu$^5$,
  P. Woudt$^3$,
  E. Tremou$^6$,
  \newauthor
  J. C. A. Miller-Jones$^7$,
  A. J. van der Horst$^{8,9}$
\\
$^1$South African Astronomical Observatory, P.O Box 9, Observatory, 7935, Cape Town, South Africa\\
$^2$Astrophysics, University of Oxford, Denys Wilkinson Building, Keble Road, Oxford OX1 3RH\\
$3$INAF--Osservatorio Astronomico di Brera, via E.\,Bianchi 46, 23807 Merate (LC), Italy\\
$^4$Department of Astronomy, University of Cape Town, Private Bag X3, Rondebosch 7701, South Africa\\
$^5$Shanghai Astronomical Observatory, Chinese Academy of Sciences,\\ 80 Nandan Road, Shanghai
200030, China \\
$^6$LESIA, Observatoire de Paris, CNRS, PSL, SU/UPD, Meudon, France \\
$^7$International Centre for Radio Astronomy Research - Curtin University, GPO Box U1987, Perth, WA 6845, Australia \\
$^8$Department of Physics, The George Washington University, 725 21st St. NW, Washington, DC 20052, USA \\
$^9$Astronomy, Physics and Statistics Institute of Sciences (APSIS), The George Washington University, Washington, DC 20052, USA
}

\date{Last updated 2015 May 22; in original form 2013 September 5}

\pubyear{2015}

\begin{document}
\label{firstpage}
\pagerange{\pageref{firstpage}--\pageref{lastpage}}
\maketitle

\begin{abstract}
We present the results of a weekly monitoring of the new black hole candidate X-ray binary MAXI J1631-472 carried out with the MeerKAT radio interferometer, the \textit{Neil Gehrels Swift Observatory}, and the Monitor of All-sky X-ray Image (\textit{MAXI}) instrument, during its 2018-2019 outburst. The source exhibits a number of X-ray states, in particular both high- and low-luminosity hard states bracketed by extended soft states. Radio flaring is observed shortly after a transition from hard/intermediate states to the soft state. This is broadly in agreement with existing empirical models, but its extended duration hints at multiple unresolved flares and/or jet-ISM interactions. In the hard state radio:X-ray plane, the source is revealed to be `radio quiet' at high luminosities, but to rejoin the `standard' track at lower luminosities, an increasingly commonly-observed pattern of behaviour.
\end{abstract}




\begingroup
\let\clearpage\relax
\endgroup
\newpage

\section{Introduction}

X-ray binaries are stellar systems which comprise a normal star (the donor), usually still undergoing nuclear fusion, and a remnant of an evolved star, the compact object, which is either a neutron star or a black hole. In these systems X-ray emission is powered by the accretion of matter from the donor to the compact object. Depending on the spectral type of the donor, X-ray binaries are generally divided up into high mass X-ray binaries (O/B spectral type; $M_{\textrm{donor}} \geq 10$M$_\odot$) and low mass X-ray binaries (spectral type later than A; $M_{\textrm{donor}} \leq 1$M$_\odot$). In low mass X-ray binaries (LMXBs), the compact object, either a neutron star (NS) or a black hole (BH), accretes matter from the donor through an accretion disc via Roche Lobe overflow. BH LMXBs are transient in nature, since they spend a majority of the time in quiescence and occasionally go into outburst where the X-ray flux is seen to increase by several orders of magnitude \citep{2006AAS...209.0705R}. These outbursts can last from weeks to years \citep{2016ASSL..440...61B}. The varying X-ray emission has an associated radio emission, and these two quantities are typically correlated in the hard spectral state \citep{2000A&A...359..251C,2003A&A...400.1007C,2001MNRAS.322...31F, 2004MNRAS.347L..52G,2012Sci...337..540F}. 

During outbursts, transient BH LMXB usually sample different states, which are defined by their X-ray spectral and timing properties, and which are connected with the properties of the radio jet \citep{2010arXiv1007.5404B}. At the outset of the outburst, when the X-ray flux is low ($L_X \sim 10^{33.5}$~erg. s$^{-1}$), the system typically lies in the hard state and the emission is associated with inverse Compton up-scattering of seed photons in a coronal region close to the accretion disc \citep{2016ASSL..440...61B}. The associated radio emission is typically weak ($L_R < 10^{28}$~erg. s$^{-1}$) with a flat spectral index. As the system goes further into an outburst, the flux rises. At a certain X-ray luminosity level - which varies in a large range from source to source and from outburst to outburst for a single source \citep{2016ApJS..222...15T} - the system transitions to an intermediate state and then enters the soft state. In the intermediate state, optically thin relativistic radio jets can be observed. As the system enters the soft spectral state, the X-ray emission is thermal, which has been attributed to an optically thick geometrically thin accretion disc, while the radio emission decreases as the jet is quenched \citep{2001ApJ...554...43C,2011MNRAS.414..677C}. These spectral state changes typically trace out a distinct path in the hardness-intensity diagram \citep{2004ApJ...611L.121Y,2007ApJ...663.1309Y,2009ApJ...701.1940Y,2011BASI...39..409B,2012Sci...337..540F}, which is associated with a transition from a compact jet to an episodic jet \citep{2004MNRAS.355.1105F}.

MAXI J1631-472 was first detected as a bright X-ray source by Monitor of All-sky X-ray Image/Gas Slit Camera (\textit{MAXI/GSC}; \citealt{2009PASJ...61..999M}) on 21 December 2018, and was initially mistaken for the X-ray pulsar AX MAXI J1631-4752 \citep{2018ATel12320....1K}. Further observations with the BAT instrument on-board the \textit{Neil Gehrels Swift Observatory} (\textit{Swift}; \citealt{2004ApJ...611.1005G}) and Nuclear Spectroscopic Telescope Array (\textit{NuSTAR}; \citealt{2010SPIE.7732E..0SH}) revealed that the position of MAXI J1631-472, with a separation of $8.4^\prime$, is not consistent with that of AX MAXI J1631-4752. MAXI J1631-472 was therefore classified as a new X-ray transient \citep{2018ATel12340....1M}. The NuSTAR spectrum (energy band $3-79$~keV) was fit with a blackbody disc temperature of $1.12\pm 0.01$~keV and a power-law index of $2.39\pm 0.02$. The spectral fitting revealed an iron K-alpha emission line with an equivalent width of 90~eV, indicating evidence of a reflection component in the spectrum \citep{2018ATel12340....1M}. The spectral properties suggested that MAXI J1631-472 is a binary system with an accreting black hole detected by \textit{NuSTAR} in the soft state. A radio counterpart was detected at the position consistent with that from the X-ray observations \citep{2019ATel12396....1R}.\\

Since 12 January 2019 (MJD~58495), MAXI J1631-472 has been monitored with MeerKAT as part of the ThunderKAT Large Survey Project \citep{2017arXiv171104132F}. In this manuscript we report on the results of the monitoring of MAXI J1631-472 with the MeerKAT interferometer, complemented with X-ray data from \textit{Swift} and \textit{MAXI}. We show the evolution of the radio and X-ray emission throughout its outburst and provide a physical interpretation of the dataset. We also compare the behaviour observed with other known outbursting BH systems. \\
The paper is structured as follows: in section~\ref{sec:obs} we describe the observations and details of the data reduction. In section~\ref{sec:results} we present the results of the two wavebands and highlight some of the interesting behaviour. In section~\ref{sec:conclusion} we summarise our conclusions. 

\section{Observations}
\label{sec:obs}
\subsection{MeerKAT}
The MeerKAT monitoring of MAXI J1631-472 was performed weekly, with the observations carried out at a central frequency of 1.28~GHz and bandwidth of 856~MHz. These observations were taken as part of the {\em ThunderKAT} Large Survey Programme on MeerKAT \citep{2016mks..confE..13F}. Each observation consists of a 15 minute scan of the target, and a two-minute scan of the phase calibrator, in this case J1726-5529. The primary calibrator used was J1939-6342, which was observed at the start of each observation for 5 minutes. We used \textsc{casa} \citep{2007ASPC..376..127M} to perform the data reduction. We performed flagging of radio frequency interference (RFI), where we removed the first and last 150 channels (channel width $\sim$ 209~kHz) of the band. The auto-flagging algorithms, \textsc{rflag} and \textsc{tfcrop}, were used for subsequent flagging. We then performed calibration, where the flux density on the primary calibrator was set using the known model. The phase-only and antenna-based delay corrections on the primary calibrator were then solved for, then the bandpass corrections for the primary were then applied. We solved for the complex gains on the primary and secondary, then proceeded to scale the gain corrections from the primary to the secondary and target source. Finally, the imaging was performed using \textsc{wsclean} \citep{2014MNRAS.444..606O}.

Flux densities for MAXI J1631-472 were extracted from the images using PyBDSF \citep{2015ascl.soft02007M}; these are presented in Table~\ref{tab:fluxes}. We were able to extract an in-band spectral index ($\alpha$, with the flux density $S_\nu$ scaling with frequency $\nu$ as $S_\nu \propto \nu^\alpha$) at the four brightest epochs using a fit to eight uniformly-spaced channels across the 856~MHz bandpass. To determine the spectral index, we performed a least-squares fit using the \textsc{scipy} and \textsc{curve-fit} module and function, respectively, in \textsc{python} to calculate the slope and its associated error bar. The results from these fits are presented in Table~\ref{tab:spec_ind} and indicate that the source was optically thin during the times of strongest radio flux density.

\begin{table}
	\centering
	\caption{MeerKAT radio (1.28 GHz) and \textit{Swift} X-ray ($1-10$~keV) fluxes of MAXI J1631-472. The spectral states of each of the observations are indicated in the final column: Soft (S), intermediate (I) and Hard (H).}
	\label{tab:RV_table}
    \setlength\tabcolsep{2pt}
	\begin{tabular}{cccc} 
		\hline\hline
		MJD & Radio flux density  & \textit{Swift} $1-10$~keV  & Spectral \\
		 & (mJy) &  X-ray flux ($10^{-10}$~erg/s/cm$^2$) & state \\		
		\hline
		58495	&	4.70	$\pm$	0.17	&	NA	& S \\
		58502	&	4.67	$\pm$	0.20	&	640	$\pm$ 64 & S	\\
		58509	&	6.88	$\pm$	0.27	&	562	$\pm$ 56 & H	\\
		58515	&	3.36	$\pm$	0.15	&	400	$\pm$ 40	& H/I \\
		58523	&	4.69	$\pm$	0.26	&	349	$\pm$ 35 & H/I	\\
		58530	&	4.86	$\pm$	0.19	&	264	$\pm$ 26 & H/I	\\
		58543	&	3.99	$\pm$	0.20	&	144	$\pm$ 14 & H/I	\\
		58560	&	15.61	$\pm$	0.50	&	95.2	$\pm$ 9.5 & S	\\
		58567	&	37.36	$\pm$	0.62	&	125.3	$\pm$ 13 & S	\\
		58574	&	22.29	$\pm$	0.53	&	84.4	$\pm$ 8.4 & S	\\
		58582	&	7.31	$\pm$	0.27	&	70.5	$\pm$ 7.1 & S	\\
		58588	&	3.48	$\pm$ 	0.23	&	63.3	$\pm$ 6.3 & S	\\
		58614	&	2.37	$\pm$	0.13	&	22.8	$\pm$ 2.3 & S	\\
		58621	&	2.41	$\pm$	0.21	&	27	$\pm$ 13 & S	\\
		58628	&	2.44	$\pm$	0.18	&	13.5	$\pm$ 1.4 & S	\\
		58634	&	2.66	$\pm$	0.18	&	4.36	$\pm$ 0.44 & I	\\
		58642	&	1.90	$\pm$	0.12	&	2.36	$\pm$ 0.24	& H\\
		58650	&	1.515	$\pm$	0.071	&	0.430	$\pm$ 0.043 & H	\\
		58658	&	1.14	$\pm$	0.14	&	0.740	$\pm$ 0.074 & H	\\
		58664	&	1.72	$\pm$	0.26	&	3.47	$\pm$ 0.035 & H	\\
		58678	&	0.62	$\pm$	0.10	&	36.7	$\pm$ 3.7	&  S\\
		58686	&	0.703	$\pm$	0.095	&	46.1	$\pm$ 4.6 & S	\\
		58691	&	0.76	$\pm$	0.11	&	49.5	$\pm$ 4.9	& S \\
		58699	&	1.15	$\pm$	0.18	&	44.6	$\pm$ 4.5 & S	\\
		58705	&	0.87	$\pm$	0.16	&	38.2	$\pm$ 3.8 & S	\\
		\hline
	\end{tabular}
	\label{tab:fluxes}
\end{table}

\begin{table}
	\centering
	\caption{Spectral indices during the strongest radio flux: $S_{\nu} \propto \nu^{\alpha}$}
    \setlength\tabcolsep{2pt}
	\begin{tabular}{cc} 
		\hline\hline
		MJD & Spectral index ($\alpha$)  \\
		\hline
		58560	&	-0.890	$\pm$	0.070 \\
		58567   &   -0.242  $\pm$   0.043 \\
        58574	&	-0.306	$\pm$	0.026 \\
        58582	&	-0.369	$\pm$	0.043 \\        
		\hline
	\end{tabular}
	\label{tab:spec_ind}
\end{table}


\subsection{MAXI}
We used X-ray data from The Monitor of All-sky X-ray Image (\textit{MAXI}/GSC; \citealt{2009PASJ...61..999M}) covering the $2-20$~keV and $2-4$~keV energy band. The daily-averaged count rates at the two energy bands were extracted from the \textit{MAXI}/GSC website \footnote{http://maxi.riken.jp/mxondem/}

\subsection{Swift}

Our target was also monitored by the \textit{X-ray Telescope} (XRT; \citealt{2005SSRv..120..165B}) on-board \textit{Swift} on a weekly basis throughout its outburst. The \textit{Swift} observations were taken (quasi-)simultaneously with the MeerKAT observing runs, as a part of a long-term monitoring of BH transients associated to the ThunderKAT Large Survey Project \citep{2016mks..confE..13F}. Therefore, we were able to measure weekly the simultaneous radio and X-ray emission from the target. 
We use the X-ray data obtained from the XRT instrument on board the \textit{Swift} Observatory to extract spectra in the 0.6--10 keV energy band through the \textit{Swift} XRT product generator online reduction pipeline \citep{Evans2007,Evans2009}. We fitted each spectrum in \textsc{XSPEC} \citep{1996ASPC..101...17A} with an absorbed power law, or a combination of a powerlaw + a disk-blackbody component, both modified by interstellar absorption. The addition of a Gaussian emission line centered at $\approx$6.4 keV was sometimes required by the fit. We initially left the Galactic neutral hydrogen absorption column parameter, $N_H$, free to vary and fixed the value to the average $N_H$ the fits returned, and then tied the $N_H$ across spectra, assuming that the same $N_H$ value applies to all of them. We find that $N_H$ = (3.36 $\pm$ 0.02)$ \times 10^{22}$\,cm$^{-2}$, see \citealt{2016A&A...594A.116H}), which is slightly higher than the $N_H$ measured in the direction of the source (i.e., $N_H$ $\approx$ 2.9$\times 10^{22}$\,cm$^{-2}$; \citealt{2018ATel12340....1M})
We finally measured the unabsorbed X-ray fluxes in the 1--10 keV energy band, which we used in Figs.~\ref{fig:lightcurves} and \ref{fig:correlation}. \\
The Burst Alert Telescope on-board \textit{Swift} (\textit{Swift}/BAT) was also used to monitor MAXI J1631-472 in the 15 -- 50 keV energy range throughout its outburst. The daily-averaged lightcurve was obtained from the the \textit{Swift} website\footnote{https://swift.gsfc.nasa.gov/results/transients/} and is shown in Fig.~\ref{fig:lightcurves}. 

\section{Results and discussion}
\label{sec:results}
\subsection{Evolution of the radio and X-ray emission}
\begin{figure}
	\includegraphics[width=1\columnwidth,angle=0]{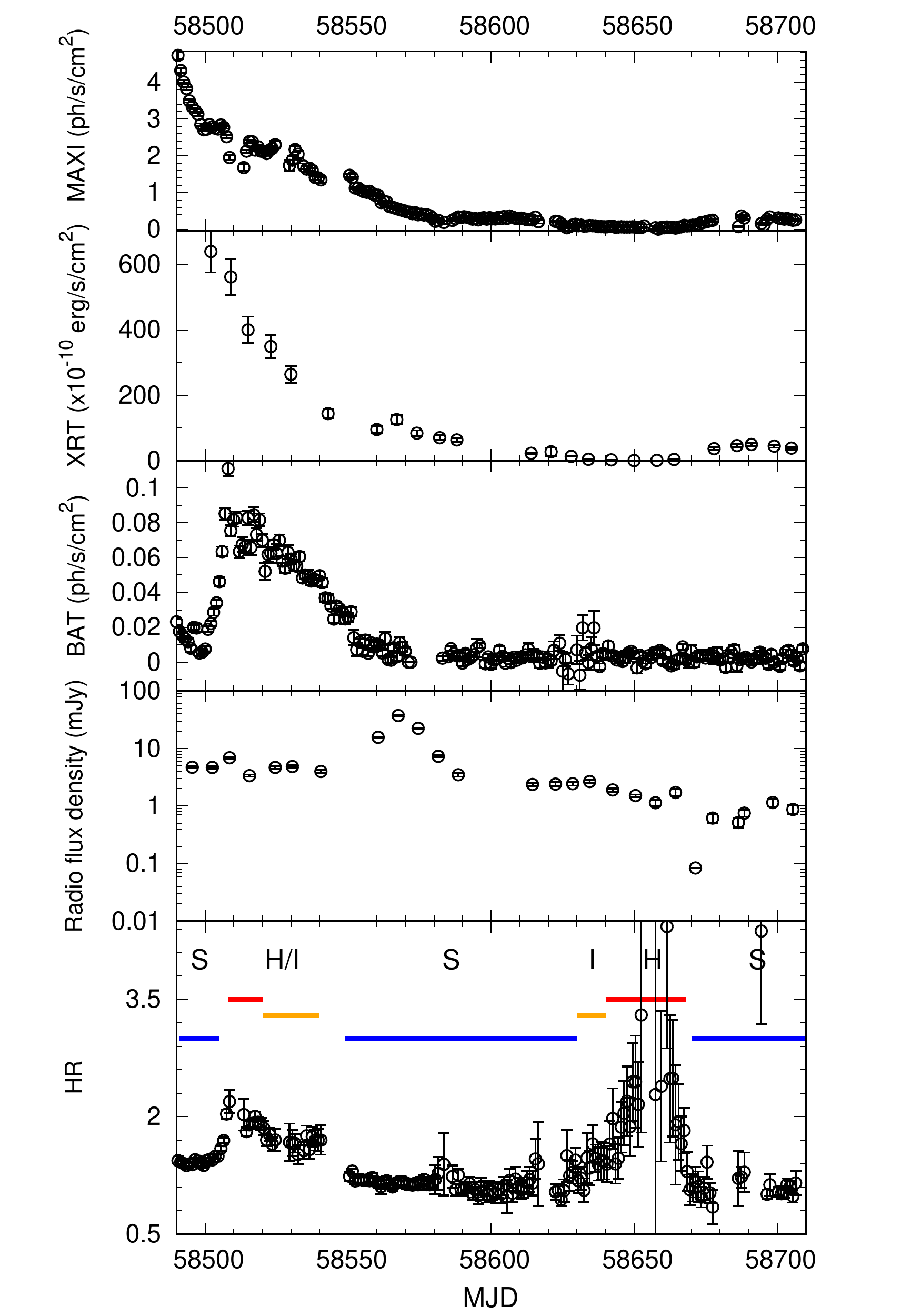}
    \caption{The evolution of the emission from MAXI J1631-472 during the 2018/19 outburst; top to bottom: \textit{MAXI} ($2-20$~keV), \textit{Swift/XRT}($1-10$~keV), \textit{Swift/BAT} ($15-50$~keV), MeerKAT (1.28~GHz), hardness ratio (HR) ($2-20/2-4$~keV) from \textit{MAXI}. The MeerKAT error bars are plotted but too small to be clearly visible on this scale. Bottom panel: The blue, orange and red lines show the soft (S), intermediate (I) and hard (H) states, respectively.} 
    \label{fig:lightcurves}
\end{figure}

In Fig.~\ref{fig:lightcurves} we show the simultaneous X-ray (\textit{MAXI} and \textit{Swift/BAT} at $2-20$~keV and $15-50$~keV, respectively) and radio light curves throughout the outburst (top three panels), and the evolution of the hardness ratio (HR, bottom panel), which is defined as the ratio of the flux in the $2-20$~keV to $2-4$~keV \textit{MAXI} energy bands. As seen in the figure, MAXI J1631-472 underwent various state transitions throughout our monitoring campaign. The radio monitoring commenced around the time of the peak of the X-ray outburst (MJD~58495), with a flux density of $\sim$4.7~mJy. The system at this stage was in the soft state, as seen in the hardness ratio (HR $\sim 1.5$). During the X-ray decline there was a change in state from the soft to hard at MJD~58508 (HR$\sim$2), with a slight increase in radio flux by $\sim$2~mJy from the previous week's measurement. This change in state was also reported by \cite{2019ATel12440....1E}.
\cite{2020MNRAS.492.3657F} use \textit{INTEGRAL/IBIS} observations during this period ($\sim$MJD~58507$-$MJD 58514) to perform spectral analysis, where they demonstrate that during the transition to the hard state the system was dominated by a hard Comptonised component with an electron temperature of $kT \sim29$~keV. As the outburst continued, MAXI J1631-472 transitioned to the intermediate state where both the soft and hard bands displayed high flux levels ($\sim$MJD~58515$-$MJD~58522; HR$\sim$1.8).
During this transition the radio flux density was back to its stable level of $\sim 4-5$~mJy.
\cite{2020ApJ...893...30X} performed spectral analysis from NuSTAR observations during a similar period ($\sim$MJD~58500$-$MJD~58515), where they demonstrated that MAXI J1631-472 transitioned from a disc-dominant state ($\sim$MJD~58500) to a power-law-dominant state ($\sim$MJD~58515).\\ 

The system then re-entered the soft state as seen in a sharp decline in the hardness ratio to HR$\sim$1.2, accompanied by a radio flare which peaked at nearly 50 mJy (between MJD~58560 and MJD~58582). MAXI J1631-472 stayed in the soft state throughout the duration of the radio flare and even after the flare had declined, with the radio flux density gradually decreasing to a flux level lower than before the flare ($\leq$3~mJy; MJD~58614). The system switched back to the hard state (MJD~58642), seen in the sharp increase in hardness ratio (HR$\sim$2.2), with the radio and X-ray flux emission continuing to gradually decline as the source went into quiescence. \\

\begin{figure*}%
    \centering
    \subfloat[ ]{{\includegraphics[width=5.25cm]{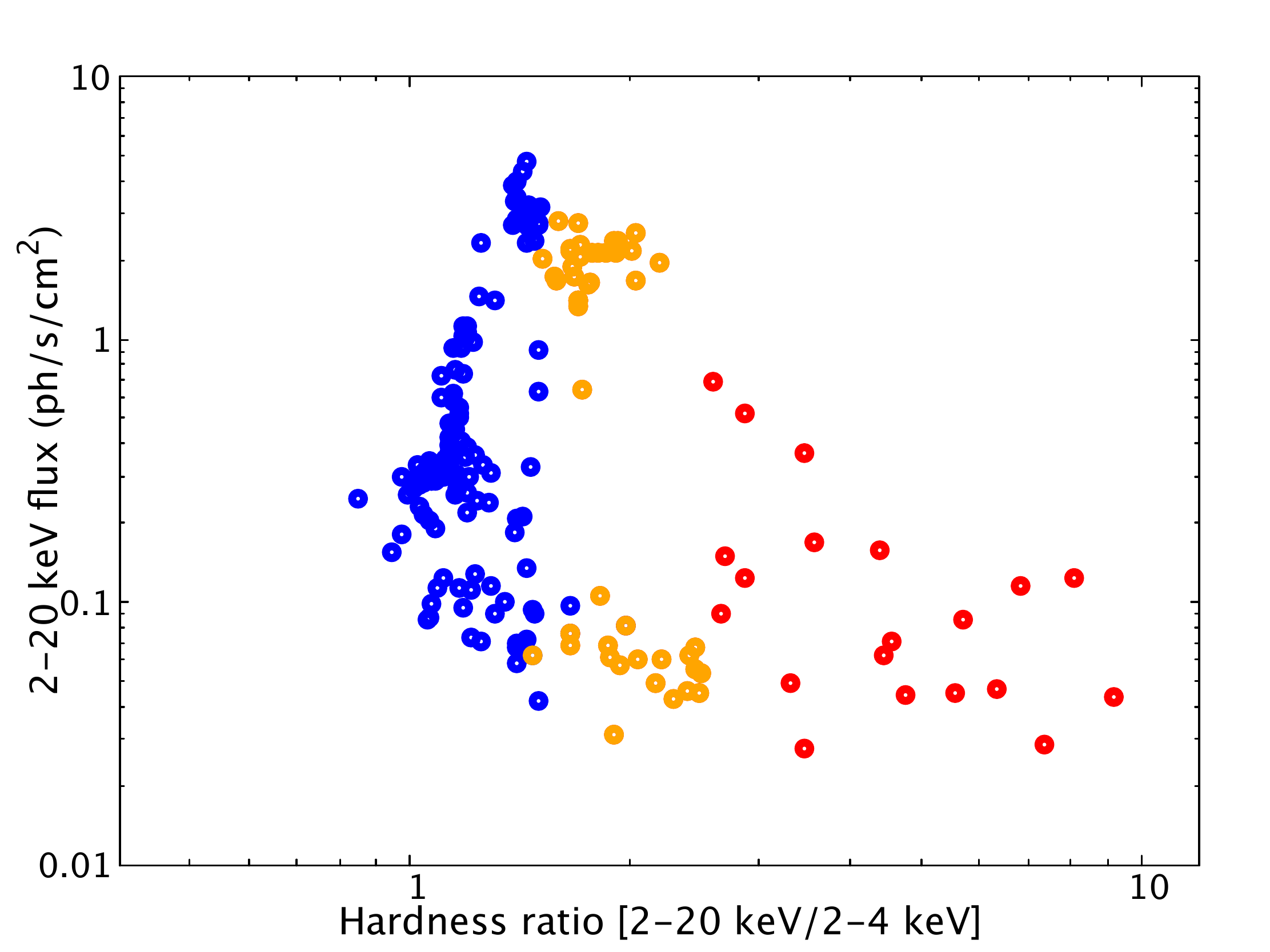}}}
    \qquad
    \subfloat[ ]{{\includegraphics[width=5.25cm]{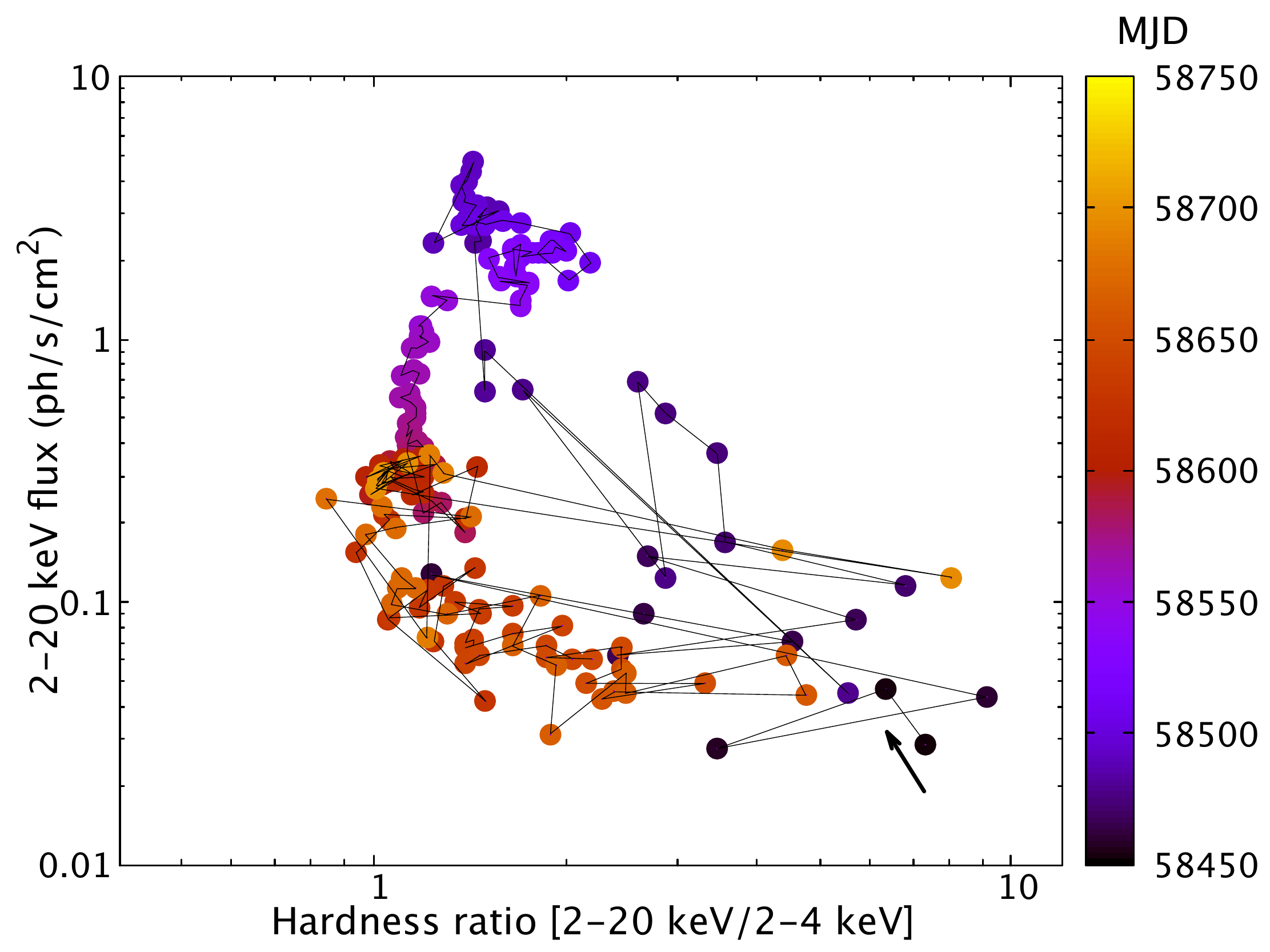} }}%
    \qquad
    \subfloat[ ]{{\includegraphics[width=5.25cm]{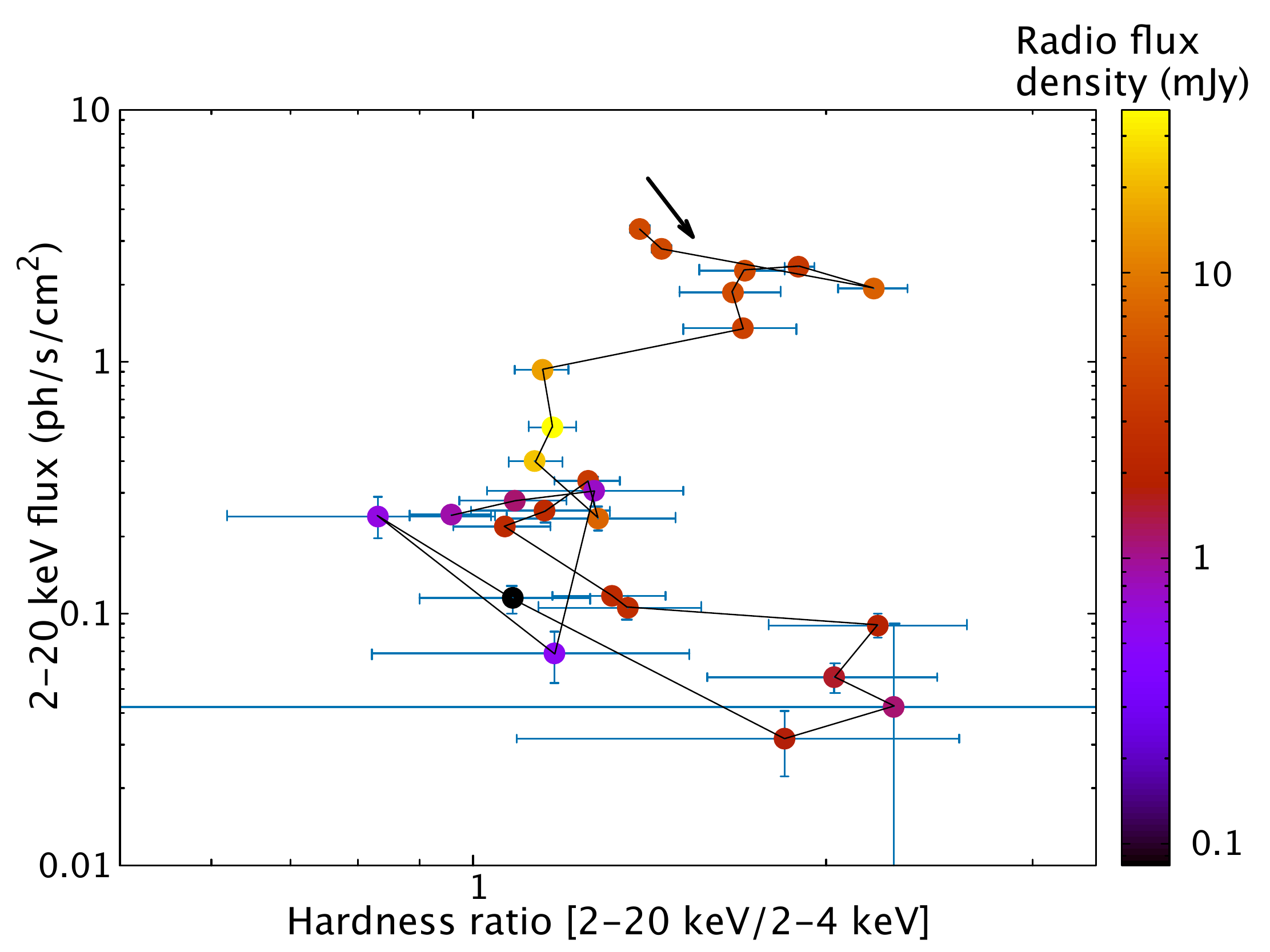} }}%

    \caption{Hardness intensity diagrams (HID) of MAXI J1631-472 during the December 2019 outburst. {\em Left, (a)}: The HID with the different spectral states colour-coded using the same criteria as the bottom panel of Fig.~\ref{fig:lightcurves} (hard=red, soft=blue, intermediate=orange).
    {\em Centre, (b)}: Full HID in which the colour bar indicates time. {\em Right, (c)}: The HID for instances when there are simultaneous radio and X-ray measurements. The colour bar in this plot indicates the corresponding radio flux density. The lines connect observations that are continuous in time. The arrow in both centre and right plots indicates the starting point of the evolution. }%
    \label{fig:HID}%
\end{figure*}

\subsection{Hardness intensity diagram}
The hardness intensity diagram (HID) for MAXI J1631-472 during the 2018/19 outburst is shown in Fig.~\ref{fig:HID}. Fig.~\ref{fig:HID} (a) shows the HID for the full duration of the outburst, with the colour bar representing the time evolution. As seen in Fig.~\ref{fig:HID} (b), MAXI J1631-472 shows a rapid rise in photon flux as it changes from quiescence to peak flux in roughly 25 days while the spectral state changes from hard (HR$\sim$9) to soft (HR$\sim$1.2) during this period. The decline of the X-ray photon flux occurs while the system is in the soft state, during which the radio flare is seen to occur (colour-bar in Fig.~\ref{fig:HID} (c))). As the X-ray photon flux continues to decline, it reaches a point ($2-20$~keV flux of $\sim$0.1 ph/s/cm$^2$) when the spectral state changes to intermediate and decays to the hard state. Fig.~\ref{fig:HID} (a) shows the HID with the colour coding indicating the different spectral states. \\

The HID track of MAXI J1631-472 follows a diagonal transition during the rise of the outburst when the spectral state changes from hard to soft and is similar to those seen in systems such as GRO J1655--40 and XTE J1550--564 (e.g. \citealt{2010MNRAS.405.1759R,2015MNRAS.451..475U}). Some systems follow a \textit{canonical} 'q'-shaped HID pattern where the rise in X-ray flux from quiescence to peak happens in the hard state and the system changes to intermediate and then to soft at roughly constant peak flux. The differences in the shapes of the HIDs have been suggested to be due to inclination angle differences, with the 'q'-shaped HIDs occurring in low inclination systems ($i\leq 60^\circ$ \citealt{2013MNRAS.432.1330M}). This suggests that the orbit of MAXI J1631-472 is possibly highly inclined. 

\subsection{Radio flaring}

The strongest radio emission from MAXI J1631-472 occurred shortly after the transition from the intermediate to the soft X-ray state. This is broadly in agreement with the unified picture put forward in \cite{2004MNRAS.355.1105F}. The three weeks or more of bright radio emission hints at multiple flaring events, since flares from X-ray binaries tend to evolve on shorter timescales (e.g. \citealt{2019MNRAS.489.4836F}), and these shorter timescale radio flares may themselves be the superposition of multiple events which can only be discriminated at mm/IR frequencies (e.g. \citealt{2017MNRAS.469.3141T}) or with Very Long Baseline Interferometry (VLBI; e.g. \citealt{2019Natur.569..374M}). There may also be a contribution from physically separated ejecta which are fading only slowly due to jet-ISM interactions leading to {\em in situ} particle acceleration (see for example \citealt{2020NatAs...4..697B}). Such emission can persist for many months, fading slowly, which is the most likely explanation for the persistent low level radio emission throughout the soft state. This component, originating in ejecta completely separated from the core may potentially contribute to the radio flux measured from the source after its return to the hard state (see further discussion in section 3.4).

Assuming that the radio flare peak was due to a transition from optically thick to thin emission as the ejecta expanded, we may use the single-frequency estimates of the physical parameters from \cite{2019MNRAS.489.4836F} (their equations 28 -- 31). For a distance of 5 kpc, at a peak flux density of 40 mJy at 1.4 GHz, we find a corresponding physical radius at peak of $R = 3 \times 10^{13}$cm, corresponding to a minimum internal energy of $E = 4 \times 10^{38}$ erg s$^{-1}$, a magnetic field of $B = \sim 0.2$~G and a resulting brightness temperature of $T_B = 5 \times 10^{10}$~K.  The dependence on distance $d$ of these estimates are different for each quantity: $R \propto d^{16/17}$, $E \propto d^{40/17}$, $B \propto d^{-4/17}$ and $T_B \propto d^{2/17}$ \citep{2019MNRAS.489.4836F}. These estimates are reasonable in the context of other X-ray binaries and rather typical of a low-mass X-ray binary (and notably considerably less powerful than the most luminous jet sources such as GRS 1915+105 or Cygnus X-3; see \citealt{2019MNRAS.489.4836F}). However, the spectral index measurements (Table~\ref{tab:spec_ind}) indicate that the source may well have been optically thin throughout the phase of brightest radio emission. In addition, as noted above, this is unlikely to have been a single event. Therefore the energy estimates remain highly uncertain.

\begin{figure*}
	\includegraphics[width=17cm,angle=0]{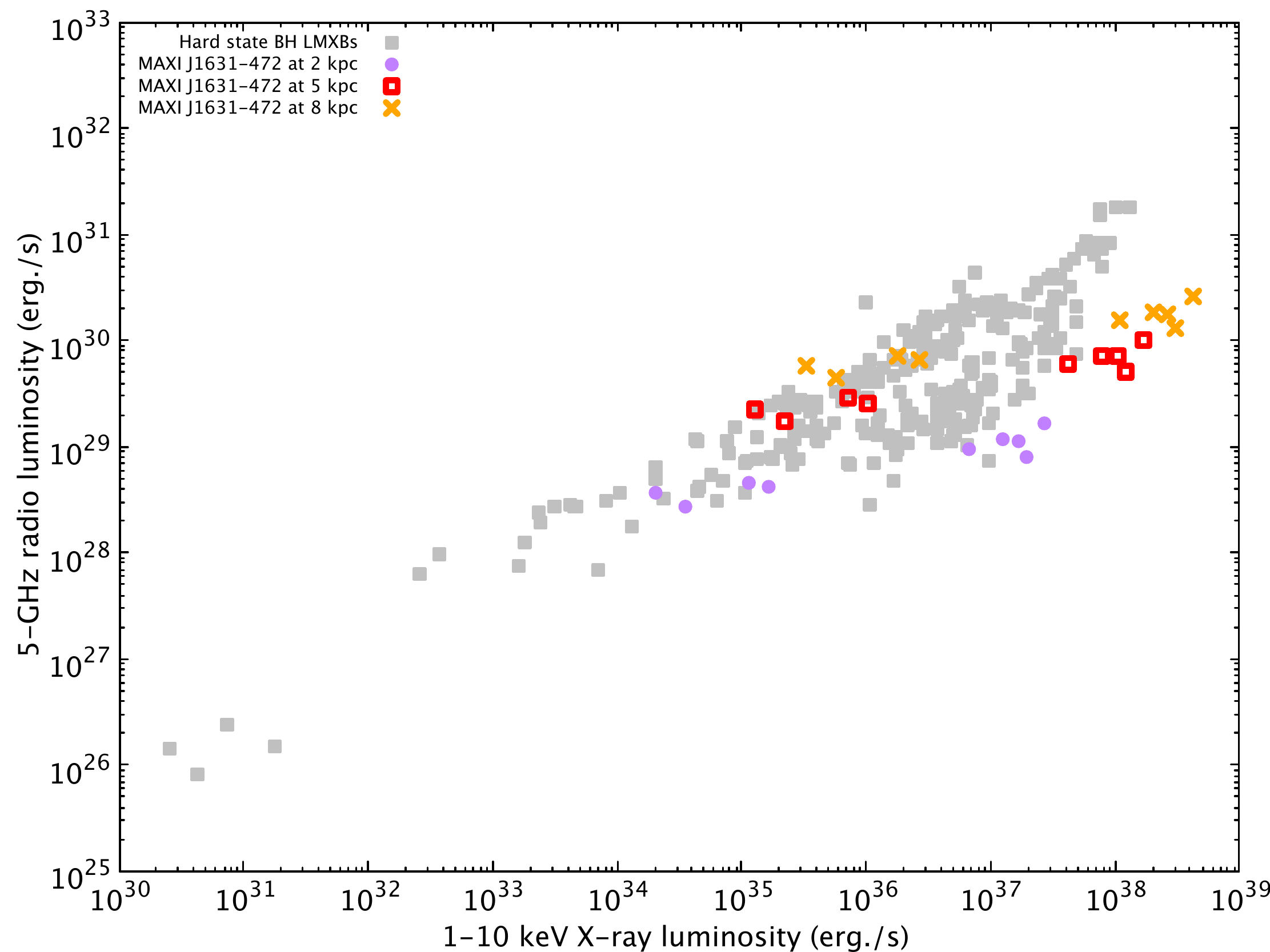}
    \caption{The radio/X-ray correlation plot which includes the hard state BH LMXBs shown with the grey filled squares from the compiled database by \citet{arash_bahramian_2018_1252036}. The hard state measurements for MAXI J1631-472 are shown using different distances: 2~kpc (purple filled circles), 5~kpc (red open squares), 8~kpc (orange crosses).} 
    \label{fig:correlation}
\end{figure*}

\subsection{Radio/X-ray correlation}

BH LMXBs in the hard state are known to show a correlation between the X-ray and radio luminosities. The `standard' correlation between the radio and X-ray emission in the black hole hard state takes a non-linear form $L_\textrm{radio} \propto L_\textrm{X-ray}^{0.5-0.7}$ \citep{2000A&A...359..251C,2003MNRAS.344...60G,2014MNRAS.445..290G,2018MNRAS.473.4122E}. A second population of BH LMXBs exists, the co-called `radio-faint' systems, which have a
smaller ratio of radio to X-ray luminosity and may have a 
steeper relationship between the radio and X-ray emission of the form $L_\textrm{radio} \propto L_\textrm{X-ray}^{1.4}$ \citep{2011MNRAS.414..677C,2014ApJ...788...52C}. Several suggestions have been made to explain the physical origin of the two tracks which include variations in the magnetic field strength of the jet \citep{2009ApJ...703L..63C}, differences in the distribution of matter in the inner accretion disc \citep{2014A&A...562A.142M} and differences in the inclination angles of the binary systems \citep{2018MNRAS.478.5159M}. \cite{2018MNRAS.478L.132G}, however, suggest, from a statistical viewpoint, that the two populations are indistinct. The BH LMXB H1743-322 has been shown to be `radio quiet' at high luminosities, but to return to the `standard' track at an X-ray luminosity between $10^{34}$--$10^{35}$ erg s$^{-1}$ \citep{2011MNRAS.414..677C}. \cite{2018MNRAS.478.5159M} have argued that this is probably a general property of all `radio quiet' sources, i.e. there is only a single radio:X-ray correlation at low X-ray luminosities.

Fig.~\ref{fig:correlation} shows the radio/X-ray correlation plot which includes measurements of known BH LMXBs using the compiled measurements in \cite{arash_bahramian_2018_1252036} and hard state measurements of MAXI J1631-472 presented in this work at 5~kpc. The 1.28~GHz radio measurements from MeerKAT were converted to 5~GHz assuming a flat spectral index for comparison with measurements presented in \cite{arash_bahramian_2018_1252036}. Note that we do not currently have any reliable distance estimates for the source, so we consider distances of 2, 5 and 8 kpc.  For the entire range of distances, it appears that the earlier hard state measurements trace out the `radio-quiet' region while the three measurements taken later during the decline of the X-ray flux seem to be more consistent `radio-loud' track.  MAXI J1631-472 therefore becomes another source for which the switch from high-luminosity `radio quiet' to lower-luminosity `standard track' evolution is observed, similar to systems such as H1743-322 \citep{2011MNRAS.414..677C}, Swift J1753.5-0127 \citep{2017ApJ...848...92P}, XTE J1752-223 \citep{2012MNRAS.423.2656R} and XTE J1659-152 \citep{2012MNRAS.423.2656R}. As noted above (section 3.3) we cannot rule out some contribution from fading ejecta, launched at the earlier hard-to-soft state transition, to the radio emission in the hard state. Examining Figure 1 in detail we can see that there was a small rise in the radio emission of approximately 12\% between two consecutive observations around the time of the transition back to the intermediate state (around MJD 58630), which likely indicates the re-activation of the compact hard state jet. In the case of MAXI J1820+070 this re-activated core jet dominates the fading emission from the jet (\citealt{2020NatAs...4..697B}), and the slight re-brightening noted above suggests this is also the case for MAXI J1631-472. We note that regardless of the physical origin, the observational phenomenon of sources returning to the `standard track' at moderately low luminosities is now well established. We cannot conclusively rule out this being due to fading ejecta in every case, but it seems unlikely.

\section{Conclusion}
\label{sec:conclusion}
We have presented radio and X-ray observations of the newly-discovered BH LMXB candidate, MAXI J1631-472, during its 2018-2019 outburst. The radio data were taken at a central frequency of 1.28~GHz with the MeerKAT interferometer as part of the ThunderKAT Large Survey Project. We used publicly available data from \textit{Swift} and \textit{MAXI} to analyse its X-ray behaviour, where we find that MAXI J1631-472 undergoes various state transitions throughout the outburst. We have investigated the quasi-simultaneous MeerKAT and \textit{Swift} observations to explore the radio/X-ray correlation. A prolonged period of radio flaring begins at the transition to the second-observed soft state, and probably originates in multiple flare-ejection events and possibly additional jet-ISM interactions.  A comparison of the contemporaneous hard-state X-ray and radio measurements of MAXI J1631-472 with those from other BH systems reveals an evolution from the 'radio quiet' to `radio loud' tracks as the X-ray luminosity decreases.

\newpage
\section*{Acknowledgements}

IMM and PAW acknowledge support from UCT and NRF. WY would like to thank Dr. Shungyu Sun for checking Swift/BAT results. WY would also like to acknowledge the support in part by the National Program on Key Research and Development Project (Grant No.
2016YFA0400804) and the National Natural Science Foundation of China (grant number 11333005 and U1838203).
\addcontentsline{toc}{section}{Acknowledgements}
\section*{Data Availability}
The data underlying this article will be shared on reasonable request to the corresponding author.




\bibliographystyle{mnras}
\bibliography{references} 

\bsp	
\label{lastpage}
\end{document}